\documentclass[conference]{IEEEtran}
\IEEEoverridecommandlockouts
\usepackage{cite}
\usepackage{amsmath,amssymb,amsfonts}
\usepackage{graphicx}
\usepackage{textcomp}
\usepackage{booktabs}
\usepackage{multirow}
\usepackage{subcaption}
\usepackage{paralist}
\usepackage{placeins}
\usepackage{url}
\usepackage{tikz}
\usepackage{float}
\usetikzlibrary{arrows.meta,positioning,calc}

\graphicspath{{figures/}}

\AtBeginDocument{%
\setlength{\abovedisplayskip}{5pt plus 1pt minus 2pt}%
\setlength{\belowdisplayskip}{5pt plus 1pt minus 2pt}%
\setlength{\abovedisplayshortskip}{3pt plus 1pt minus 1pt}%
\setlength{\belowdisplayshortskip}{3pt plus 1pt minus 1pt}}
\begin{document}
\flushbottom

\title{QoI-Aware Provisional Rollout and Retrospective Reconciliation for Reduced-State Scientific Twins\thanks{The authors acknowledge the Johns Hopkins Turbulence Database for providing access to the isotropic-turbulence data used in this study. The work of Anand Rangarajan, Scott Klasky and Sanjay Ranka was supported in part by UT-Battelle, LLC, under contract DE-AC05-00OR22725 with the U.S. Department of Energy (DOE).}
}

\author{
\IEEEauthorblockN{Liangji Zhu\IEEEauthorrefmark{1}, Scott Klasky\IEEEauthorrefmark{2},
 Jaemoon Lee\IEEEauthorrefmark{2},
  Qian Gong\IEEEauthorrefmark{2},
  Anand Rangarajan\IEEEauthorrefmark{1}, and Sanjay Ranka\IEEEauthorrefmark{1}}
\IEEEauthorblockA{\IEEEauthorrefmark{1}Department of Computer \& Information Science \& Engineering, University of Florida, Gainesville, FL, USA \\
Email: zhu.liangji@ufl.edu, anand@cise.ufl.edu, ranka@cise.ufl.edu}
\IEEEauthorblockA{\IEEEauthorrefmark{2}Oak Ridge National Laboratory, Oak Ridge, TN, USA \\
Email: klasky@ornl.gov,
leej8@ornl.gov,
gongq@ornl.gov}
}

\maketitle

\begin{abstract}
Scientific twins may need to continue operating when updates from an authoritative primary system are temporarily unavailable. Once synchronization resumes, the new boundary can also be used to revise the intervening history. We distinguish an immediately available causal \emph{provisional} trajectory from a delayed, future-conditioned \emph{reconciled} trajectory.
For reduced-state twins, we introduce a deterministic, calibration-based reconciliation method. A smooth temporal bridge carries the residual observed at the next synchronization block backward through the provisional interval. An analytic energy-matching stage then applies smooth regional gains and a global rescaling to match a component-energy trajectory estimated by cubic regression in log-energy space from synchronized frames on both sides of the gap. The method uses no additional correction network and revises decoded history without changing the latent state used for later rollouts.
We evaluate 64 spatial patches from 16 JHTDB isotropic-turbulence slices for both velocity components and gaps $S\in\{4,6,8\}$. During the longest gap, field error and gradient-sensitive QoI error degrade at markedly different rates, so field error alone does not characterize provisional fidelity. At $S=8$, full reconciliation reduces window-averaged NRMSE by about 60\% for both components and global gradient-intensity error from 4.21\% to 2.91\% for $v_x$, whereas future-aware physical interpolation reaches 20.40\% on the same metric. Energy matching additionally makes the reconciled history match its boundary-inferred global energy trajectory exactly. Future boundary information therefore substantially improves scientifically relevant properties within the evaluated regime.
\end{abstract}


\section{Introduction}
\label{sec:intro}
Scientific twins couple an evolving primary system---a large-scale simulation, experimental facility, or other authoritative source of state whose output streams routinely outpace storage, transmission, and availability---to a computational model that is updated from primary data~\cite{glaessgen2012digital,rasheed2020digital}. Periodic synchronization raises a practical question: what should the secondary model provide while primary states are unavailable, and what should happen to that provisional history after synchronization resumes?

These are different reconstruction tasks. During the gap, the surrogate has only past primary-derived states and must advance \emph{causally}. It produces a provisional trajectory that is immediately available but increasingly affected by rollout drift. Once the next synchronization block arrives, the surrogate has information from both sides of the gap and can revise the decoded fields within it. This reconciled trajectory arrives later but may be more faithful. Online consumers can act only on provisional states; buffered or archival consumers can replace them with the reconciled version.

We study this paired task using an existing reduced-state surrogate. The causal predictor advances a learned latent state, whereas reconciliation operates on decoded physical fields. Our focus is trajectory fidelity rather than compression or deployment cost. We ask how scientifically important quantities change during a finite synchronization gap and how much the later primary boundary can improve the decoded history.

This distinction matters because field-level and derived-quantity errors can exhibit different position-wise degradation. On JHTDB isotropic turbulence~\cite{jhtdb}, with the CAESAR-based~\cite{caesar} surrogate used here, gradient-sensitive QoI error grows at a markedly different rate from decoded field error within a gap (Section~\ref{sec:rq1}), so NRMSE alone does not characterize component-energy and gradient-intensity fidelity, and evaluating a twin only by NRMSE can omit information needed for downstream analysis.

Rather than discard the causal rollout after synchronization returns, we retain its evolving structure and use the newly observed discrepancy to correct its recent history. A closed-form temporal bridge carries the boundary residual backward through the provisional interval. An analytic \emph{QoI-aware energy-matching correction} then controls component energy. It estimates the desired energy trajectory from synchronized boundary states in log space, applies smooth regional gains, and uses a final global rescaling to match the global target exactly. This separates temporal drift correction from amplitude correction. The parameters are chosen from calibration data, and test-time inference needs no reference fields from inside the gap or additional correction network. The result is a deterministic, block-delayed revision of decoded history; later rollouts begin from newly synchronized latent states.

Our experiments emulate a temporary loss of \emph{access} to authoritative states by withholding 4--8 frames from the surrogate while the primary trajectory continues and later becomes available again. This covers interruptions in observation, communication, output delivery, or synchronization. Retrospective reconciliation is possible only if an authoritative trailing block eventually arrives.

Our contributions are:
\begin{compactitem}
\item \textbf{A two-product formulation for block-synchronized scientific twins.} We distinguish causal provisional trajectories from future-conditioned reconciled trajectories and make their information access, latency, provenance, and uses explicit. We also separate revision of decoded history from evolution of the latent state.
\item \textbf{A representation-consistent, calibration-based reconciliation method.} After synchronization resumes, a closed-form temporal bridge uses residuals from the trailing primary-derived block to correct the provisional trajectory smoothly. Analytic energy matching then enforces a boundary-inferred global energy trajectory without using hidden primary fields from the repaired interval at test time or training another correction network. Field and gradient improvements are measured empirically rather than built into the constraint.
\item \textbf{A controlled study of degradation and recovery.} On JHTDB turbulence, we show that decoded field error alone does not characterize QoI drift. We then measure how recovery depends on position within the repaired interval and on synchronization gaps $S \in \{4,6,8\}$ for both velocity components.
\end{compactitem}

The remainder of the paper reviews related work (Section~\ref{sec:related}), defines the operating setting and QoIs (Section~\ref{sec:formulation}), presents the provisional and retrospective stages (Section~\ref{sec:method}), reports the controlled evaluation (Section~\ref{sec:experiments}), and concludes in Section~\ref{sec:conclusion}.

\section{Related Work}
\label{sec:related}

\subsection{Scientific Twins and Reduced-State Surrogates}
Scientific-twin formulations describe an evolving computational counterpart that is updated from the primary system and supports monitoring or prediction~\cite{glaessgen2012digital,rasheed2020digital}. We study one operating condition within that broader architecture: a learned surrogate is synchronized in blocks, continues causally across a finite data gap, and later revises its decoded history. We do not evaluate a complete deployed twin or its decision loop.

Neural surrogates for spatiotemporal systems include convolutional predictors, Fourier neural operators~\cite{fno}, and message-passing PDE solvers~\cite{brandstetter2022message}. Learned representations can reduce the dimensionality of the state advanced by such predictors. CAESAR~\cite{li2025foundation,caesar}, used here as the representation backbone, maps full scientific fields to latent spatial states through a variational encoder--decoder. We use that reduced state to study temporal fidelity across synchronization gaps.

\subsection{Autoregressive Continuity and Temporal Reconstruction}
Autoregressive prediction provides causal continuity, but its errors accumulate during rollout. Existing remedies act mainly forward in time, including rollout-aware training~\cite{brandstetter2022message}, training-noise injection for turbulence models~\cite{stachenfeld2022learned}, and iterative refinement of newly predicted states~\cite{lippe2023pderefiner}. These methods improve a state before it is emitted. Our method instead revises a finite provisional history after a later primary boundary becomes available.

Future-aware temporal reconstruction also estimates intermediate states from later information. CAESAR-D~\cite{li2025generative,caesar}, for example, uses latent diffusion to reconstruct scientific states from retained keyframes, while physical interpolation gives a deterministic endpoint-based estimate. Our operating path is different: the twin first emits a causal trajectory, then preserves its evolving structure while smoothing the residual observed at resynchronization.

\subsection{Finite-Interval Smoothing and Adjacent Systems Axes}
Data assimilation updates a model state as observations arrive; nudging and ensemble Kalman methods are standard examples~\cite{asch2016data}. Fixed-interval smoothers, such as the Rauch--Tung--Striebel smoother~\cite{rauch1965maximum}, use future observations to improve earlier estimates under an explicit state-space and noise model. Our bridge is a deterministic smoothing construction. It assumes that the correction varies smoothly in time, uses the residual observed at the trailing synchronization block, and reduces to a precomputable linear operator. It does not estimate uncertainty, infer process noise, or produce a Bayesian posterior. We do not benchmark against ensemble or RTS-style smoothers for a structural reason: they require a stochastic state-space model with noise statistics---and, for ensembles, many forward integrations---none of which exist in this deterministic frozen-representation setting; supplying one would constitute a different method rather than a baseline. Learned latent state-space models such as the Kalman VAE~\cite{fraccaro2017kvae} instead support smoothing within a probabilistic latent space; here the latent representation is frozen and deterministic, so we smooth decoded fields and leave latent-space reconciliation to future work.

Compression (e.g., SZ~\cite{SZ_3}, CAESAR~\cite{caesar}) addresses a separate systems axis: it reduces the bits used to retain existing fields and may transport synchronization states, but bit rate is neither the objective nor the evaluation criterion here.

\section{Problem Formulation}
\label{sec:formulation}

\subsection{Setting and State Provenance}
Let $X^{\mathrm{P}}_t \in \mathbb{R}^N$ denote the authoritative physical state of an evolving primary system and the reference used for evaluation. The surrogate receives primary-derived latent states only in synchronization blocks. In the evaluated schedule, block $j$ contains $K = 4$ consecutive synchronized latent states ending at $\tau_j$; access is then withheld for an interval
\begin{equation}
\mathcal{I}_j = \{\tau_j + 1, \ldots, \tau_j + S\},
\label{eq:interval}
\end{equation}
after which another $K$-state primary block becomes available. The sequence is therefore block-synchronized rather than uniformly sampled. The four-state block is imposed by the predictor's context length, and we do not claim that it is an optimal synchronization schedule. For $S \in \{4,6,8\}$, the nominal long-run synchronization fractions $K/(K+S)$ are $1/2$, $2/5$, and $1/3$, respectively.

The experiment emulates temporary loss of access by hiding the primary states in $\mathcal{I}_j$ from the surrogate while retaining them only for evaluation. Retrospective repair requires a later authoritative boundary: either the primary trajectory continues or the missing evolution is recomputed before resynchronization. If no later primary block becomes available, the surrogate can continue to produce provisional estimates, but it cannot reconcile them.

Reconciliation targets settings where the interior states are permanently unrecorded even though boundary blocks survive: instrument downtime between observation blocks, intermediates never stored under reduced temporal archiving, or simulation output written only in periodic blocks. If the missed states are merely delayed---for example, buffered at the source during a communication outage---they should replace the provisional interval directly once they arrive, and reconciliation is unnecessary.

We distinguish the authoritative reference $X^{\mathrm{P}}_t$ from the frozen-VAE reconstruction
\begin{equation}
X^{\mathrm{VAE}}_t = \mathcal{D}(z_t), \qquad z_t = \mathcal{E}(X^{\mathrm{P}}_t).
\label{eq:vae}
\end{equation}
The synchronization interface retains $z_t$, not the corresponding full-resolution $X^{\mathrm{P}}_t$. Consequently, $X^{\mathrm{VAE}}_t$ is the representation-consistent physical boundary available to temporal reconciliation after the synchronized latent state is decoded. Throughout, \emph{authoritative} describes the provenance of $X^{\mathrm{P}}_t$, not zero encoder--decoder error. The VAE residual $X^{\mathrm{P}}_t - X^{\mathrm{VAE}}_t$ is excluded from the repair target and synchronization residual by construction, but it remains in every reported error because evaluation uses $X^{\mathrm{P}}_t$.

\subsection{Reduced State and Two Trajectory Products}
A frozen CAESAR variational autoencoder with encoder $\mathcal{E}$ and decoder $\mathcal{D}$ maps each synchronized primary field to a reduced state $z_t \in \mathbb{R}^d$, with $d < N$; decoding that state yields $X^{\mathrm{VAE}}_t$. A latent predictor $\mathcal{P}$ maps four consecutive latent states to the next four,
\begin{equation}
(\hat{z}_{t+1}, \ldots, \hat{z}_{t+4}) = \mathcal{P}(z_{t-3}, \ldots, z_t),
\label{eq:predictor}
\end{equation}
and is applied recursively over $\mathcal{I}_j$. The decoded causal estimate
\begin{equation}
\hat{X}^{-}_t = \mathcal{D}(\hat{z}_t), \qquad t \in \mathcal{I}_j,
\label{eq:provisional}
\end{equation}
is the \emph{provisional} trajectory. It is available without waiting for the next synchronization block.

After that block arrives, its residual relative to the twin rollout is used to construct a retrospective correction $\mathcal{R}_{\ell}$ for position $\ell = t - \tau_j$:
\begin{equation}
\hat{X}^{+}_t = \hat{X}^{-}_t + \mathcal{R}_{\ell}(R_{\mathrm{sync}}).
\label{eq:reconciled}
\end{equation}
The revised field $\hat{X}^{+}_t$ belongs to the \emph{reconciled} trajectory. It becomes available only after resynchronization and cannot change actions already taken from $\hat{X}^{-}_t$. Reconciliation acts only on decoded physical fields. It neither produces corrected latent states nor alters the state used for the next autonomous rollout. For this reason, we use \emph{trajectory reconciliation} rather than twin-state assimilation.

For a QoI map $q(\cdot)$ and authoritative reference $X^{\mathrm{P}}_t$, we distinguish provisional and reconciled errors,
\begin{equation}
e^{-}_{q,t} = \mathrm{err}\big(q(\hat{X}^{-}_t), q(X^{\mathrm{P}}_t)\big), \quad
e^{+}_{q,t} = \mathrm{err}\big(q(\hat{X}^{+}_t), q(X^{\mathrm{P}}_t)\big).
\label{eq:errors}
\end{equation}
The field-error decomposition makes the scope of repair explicit:
\begin{align}
X^{\mathrm{P}}_t - \hat{X}^{-}_t &= (X^{\mathrm{P}}_t - X^{\mathrm{VAE}}_t) + (X^{\mathrm{VAE}}_t - \hat{X}^{-}_t), \label{eq:decomp1}\\
X^{\mathrm{P}}_t - \hat{X}^{+}_t &= (X^{\mathrm{P}}_t - X^{\mathrm{VAE}}_t) + (X^{\mathrm{VAE}}_t - \hat{X}^{-}_t - \mathcal{R}_{\ell}). \label{eq:decomp2}
\end{align}
The temporal smoother is fitted to reduce the second term: temporal prediction drift relative to the fixed representation. It neither observes nor targets the first term, which is VAE reconstruction error. We ask how $e^{-}_{q,t}$ changes with position in the provisional interval, how much resynchronization reduces it, and how both depend on $S$. Because the evaluation does not define an operational error threshold, we report error versus interval position and gap length rather than claim a certified safe or trustworthy horizon.

\subsection{Physics-Based Quantities of Interest}
\label{sec:qoi}
JHTDB isotropic turbulence contains large-scale energy sustained by low-wavenumber forcing and small-scale velocity gradients associated with dissipation. We define each QoI separately for $v_i$, $i \in \{x, y\}$.

\textbf{Component-wise velocity energy}, $E_i = \tfrac{1}{2} v_i^2$, measures the preservation of each component's intensity. Its global and regional forms are
\begin{align}
E^{G}_{i} &= \frac{1}{HW} \sum_{x,y} E_i(x,y), \label{eq:eg}\\
E^{R}_{i,mn} &= \frac{1}{|\Omega_{mn}|} \sum_{(x,y) \in \Omega_{mn}} E_i(x,y), \label{eq:er}
\end{align}
where the field is divided into non-overlapping $16 \times 16$ blocks $\Omega_{mn}$ and each regional QoI is the block-wise spatial average.

\textbf{Component-wise longitudinal gradient intensity}, $D_i = (\partial v_i / \partial i)^2$, measures small-scale velocity variation. Under isotropy, its \emph{global mean} is proportional to the mean energy-dissipation rate, so $D^{G}_{i}$ is a useful proxy for dissipative activity. The regional maps $D^{R}_{i,mn}$ describe how gradient intensity is distributed in space; they are not local dissipation estimates. We define $D^{G}_{i}$ and $D^{R}_{i,mn}$ in the same way as the corresponding energy quantities.

Spatial derivatives are computed on interior grid points using the second-order centered stencil $(v_{j+1}-v_{j-1})/(2\Delta)$, with no periodic wrap-around. We use unit index spacing, $\Delta=1$; the corresponding constant physical grid-spacing factor cancels in the reported relative $D_i$ errors. Thus, $D_x=(\partial v_x/\partial x)^2$ and $D_y=(\partial v_y/\partial y)^2$.
For a $256\times256$ patch, the centered derivative has support $254\times256$ for $D_x$ and $256\times254$ for $D_y$. Regional gradient-intensity QoIs are computed over non-overlapping $16\times16$-pixel blocks; only complete blocks are retained, with the high-index remainder excluded before pooling. This yields regional maps of size $15\times16$ for $D_x$ and $16\times15$ for $D_y$.

\begin{figure}[!t]
\centering
\includegraphics[width=0.70\columnwidth]{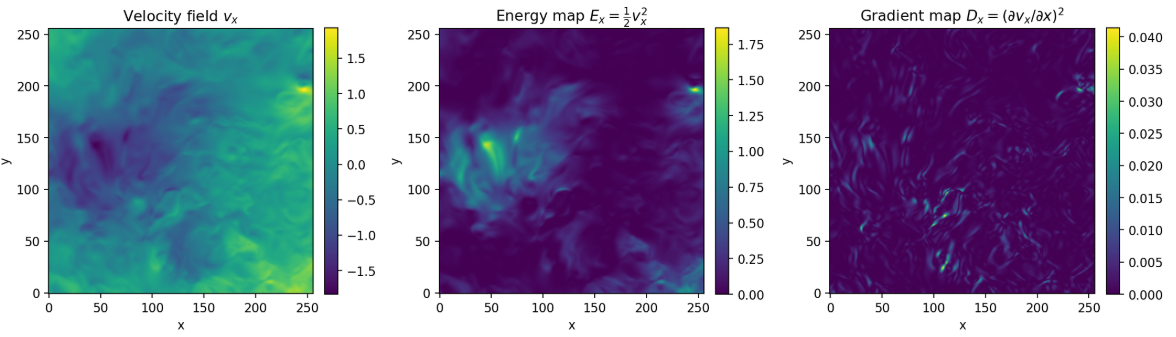}
\caption{Velocity field $v_x$ (left), energy map $E_x = \tfrac{1}{2}v_x^2$ (center), and gradient map $D_x = (\partial v_x / \partial x)^2$ (right) for a representative JHTDB frame.}
\label{fig:qoimaps}
\end{figure}

Global QoI fidelity is evaluated by relative error of the global statistic; regional fidelity by the relative $L_2$ error between predicted and reference regional maps. Fig.~\ref{fig:qoimaps} illustrates the three fields for a representative frame: the velocity field, its energy map, and its gradient map, which concentrates on fine-scale filamentary structures that aggregate metrics do not resolve.

\section{Methodology}
\label{sec:method}

\begin{figure*}[!t]
\centering
\includegraphics[width=0.75\textwidth]{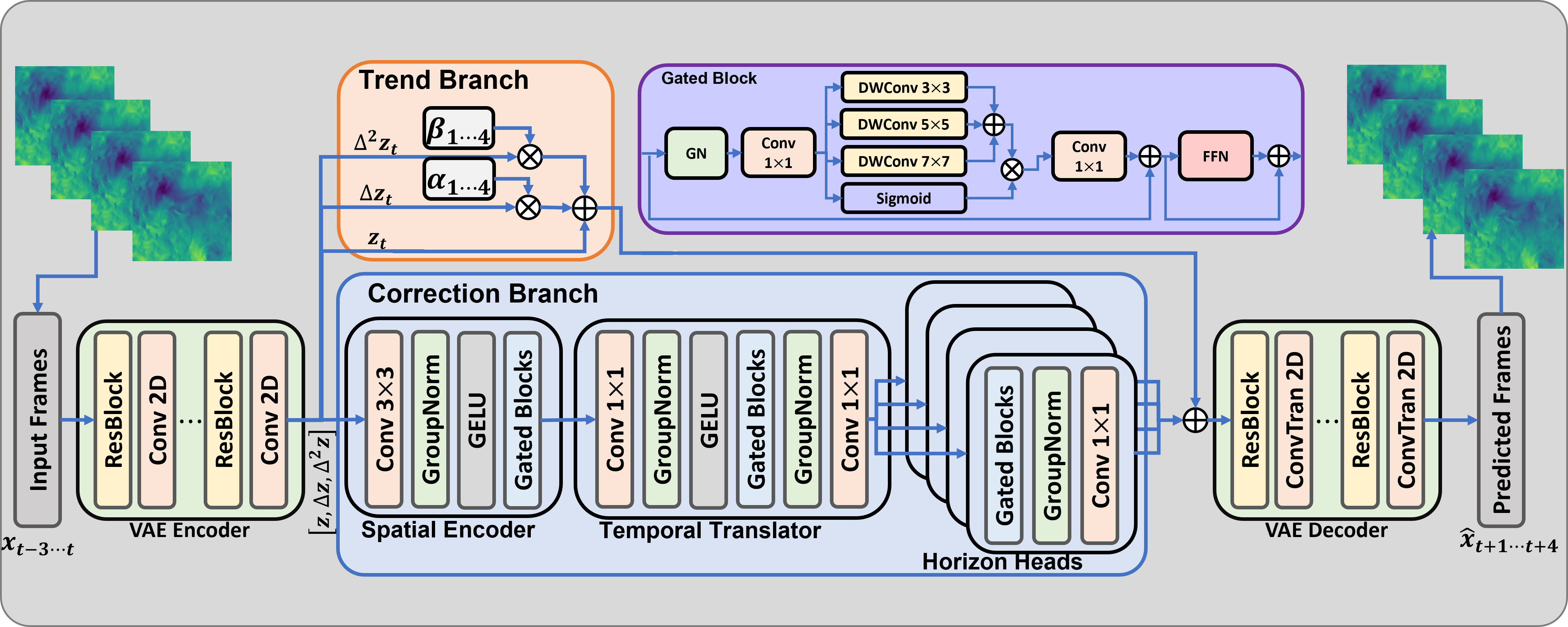}
\caption{End-to-end architecture. Four input frames are encoded independently by the frozen VAE encoder into latent states. The latent history and its first- and second-order temporal differences $(\Delta z_t, \Delta^2 z_t)$ feed two paths: a trend branch with learned per-position, per-channel coefficients $\alpha_{1\ldots4}$ and $\beta_{1\ldots4}$, and a correction branch comprising a spatial encoder, a temporal translator built from gated convolutional blocks, and four horizon-specific heads that produce the learned residuals. Their sum yields the next four latent states, which the frozen VAE decoder maps back to physical fields.}
\label{fig:model}
\end{figure*}

\subsection{Causal Provisional Rollout}

\textbf{Latent representation.}
The representation model consists of the encoder $\mathcal{E}$ and decoder $\mathcal{D}$
applied independently to each temporal state. For each single-component velocity field $v_t\in\mathbb{R}^{256\times256}$, the encoder produces
$z_t=\mathcal{E}(v_t)\in\mathbb{R}^{64\times16\times16}$, and the decoder maps $z_t$ back to the original spatial resolution.

The encoder contains four residual stages with channel widths $16$, $32$, $48$, and $64$. Each stage is followed by a factor-two spatial downsampling operation, yielding a final latent spatial resolution of $16\times16$. The decoder uses the reverse channel hierarchy with four corresponding upsampling stages to reconstruct the $256\times256$ field. The encoder--decoder contains $338{,}049$ parameters and is trained separately from the temporal predictor. Its parameters are frozen before predictor training.

\textbf{Latent predictor.} The predictor follows a SimVP-style encoder--translator design~\cite{simvp}, with gated convolutional blocks inspired by SimVPv2~\cite{simvpv2}. Given four latent states $\{z_{t-3},z_{t-2},z_{t-1},z_t\}$, it jointly predicts the next four states. The latent history and its temporal-difference features are mapped to a hidden width of $192$, processed by three shared encoder blocks and a six-block temporal translator, and decoded by four horizon-specific two-block heads. For forecast position $\ell\in\{1,\ldots,4\}$,
\begin{equation}
\hat z_{t+\ell}=z_t+\alpha_{\ell}\Delta z_t+\beta_{\ell}\Delta^2 z_t+r_{\ell},
\end{equation}
where $\Delta z_t=z_t-z_{t-1}$, $\Delta^2 z_t=z_t-2z_{t-1}+z_{t-2}$, and $r_{\ell}$ is the learned residual. The coefficients $\alpha_{\ell}$ and $\beta_{\ell}$ are learned per forecast position and latent channel, with $\beta_{\ell}$ bounded by a hyperbolic tangent. Longer rollouts recursively feed each predicted four-state block back to the same predictor.

The information flow is strictly layered: synchronized primary fields enter only through $\mathcal{E}$, every physical output leaves only through $\mathcal{D}$, and between them the predictor is the sole trained temporal component, consuming the four most recent latent states---retained synchronized states at the start of a gap, its own fed-back predictions thereafter. Reconciliation sits entirely downstream of the decoder: it revises decoded history in physical space and never feeds a repaired state back into the predictor or the latent sequence.

This stage is strictly causal. It does not use the length or contents of the future synchronization block, and it remains the only available output during loss of synchronization. Its empirical QoI drift is evaluated in Section~\ref{sec:rq1}.

\subsection{Boundary-Anchored Temporal Residual Smoothing}
\label{sec:bridge}
When primary-derived latent states resume, their frozen-VAE reconstructions reveal how far the predictor has drifted at the trailing boundary. Let $\hat{X}_{\mathrm{sync}} \in \mathbb{R}^{K \times N}$ be the decoded rollout at the $K$ trailing synchronization positions, with each physical field flattened. Let $X^{\mathrm{VAE}}_{\mathrm{sync}}$ be the frozen-VAE reconstructions obtained by decoding the synchronized latents at the same positions. Their difference is the synchronization residual,
\begin{equation}
R_{\mathrm{sync}} = X^{\mathrm{VAE}}_{\mathrm{sync}} - \hat{X}_{\mathrm{sync}} \in \mathbb{R}^{K \times N}.
\label{eq:rsync}
\end{equation}
Both terms in $R_{\mathrm{sync}}$ are decoded fields from the same fixed VAE representation. The residual therefore measures temporal prediction drift relative to a representation-consistent boundary. It excludes the VAE reconstruction residual $X^{\mathrm{P}}_{\mathrm{sync}} - X^{\mathrm{VAE}}_{\mathrm{sync}}$. We infer a correction $C \in \mathbb{R}^{L \times N}$ for the preceding $L$ provisional fields by solving
\begin{equation}
\min_{C} \; \|C\|_F^2 \;+\; \lambda \left\| D_2 \begin{bmatrix} 0 \\ 0 \\ C \\ R_{\mathrm{sync}} \end{bmatrix} \right\|_F^2,
\label{eq:bridgeobj}
\end{equation}
where $D_2$ is the second-order temporal difference operator, $(D_2 z)_t = z_t - 2 z_{t+1} + z_{t+2}$, applied along the temporal axis of the stacked sequence.

The term $\|C\|_F^2$ discourages unnecessary changes to the provisional trajectory, while the second term penalizes temporal curvature in the stacked correction sequence. The two leading zeros represent the preceding primary-derived history and softly encourage a near-zero incoming slope; the trailing residual block softly encourages a smooth transition toward the discrepancy observed at resynchronization. Neither constraint is imposed exactly.
In all reported experiments, the repaired interval spans the full synchronization gap, so $L=S$.

Because \eqref{eq:bridgeobj} is an unconstrained quadratic in $C$, it has a closed-form solution. Split the difference operator into its action $D_p$ on the unknown rows and $D_s$ on the known boundary rows:
\begin{equation}
C = \left( I + \lambda D_p^{\top} D_p \right)^{-1} \left( -\lambda D_p^{\top} D_s \right) R_{\mathrm{sync}} \;=\; W_{\lambda} R_{\mathrm{sync}},
\label{eq:closedform}
\end{equation}
Here, $W_{\lambda} \in \mathbb{R}^{L \times K}$ maps the $K$ observed boundary residuals to corrections for the preceding $L$ provisional fields. It depends only on $(L,K,\lambda)$ and can be precomputed for each repair geometry. Applying the bridge is therefore a temporal matrix product over the field entries. The smoother is deterministic, requires no learned weights, and uses no reference fields from the reconciled interval at test-time inference.

\subsection{QoI-Aware Energy Matching and Validation Selection}
\label{sec:energyproj}
The temporal smoother reduces trajectory drift but does not explicitly control a physical QoI. We therefore add a lightweight, \emph{analytic} amplitude correction for component energy. This operator directly targets energy; changes in field or gradient error are empirical consequences, not enforced properties. We fix all calibration choices using training and held-out validation data before test inference. Repair targets are defined using frozen-VAE reconstructions rather than the original physical fields.

\textbf{Regularization and strength selection.} Multiple rollout windows are sampled from the calibration interval. For each candidate $\lambda$, we measure how much the inferred correction explains the known representation-consistent temporal residual
\begin{equation}
R_{\mathrm{roll},l} = X^{\mathrm{VAE}}_l - \hat{X}^{-}_l.
\label{eq:rroll}
\end{equation}
For calibration window $w$, let $\widetilde{C}_{w,l} = [W_{\lambda} R_{\mathrm{sync},w}]_l$ be the smoothed correction at rollout position $l$. Its calibrated strength is
\begin{equation}
\alpha_l(\lambda) = \mathrm{clip}_{[0,1]} \frac{\sum_{w \in \mathcal{C}} \langle \widetilde{C}_{w,l}(\lambda),\, R_{\mathrm{roll},w,l} \rangle}{\sum_{w \in \mathcal{C}} \| \widetilde{C}_{w,l}(\lambda) \|_2^2 + \epsilon},
\label{eq:alpha}
\end{equation}
where $\mathcal{C}$ is the set of calibration windows and the inner product and norm cover all spatial entries of the field. Equation~\eqref{eq:alpha} is the least-squares projection of the representation-consistent residual onto the proposed correction, clipped to $[0,1]$. We select the regularization by expected log residual reduction,
\begin{equation}
\lambda^{*} = \operatorname*{argmin}_{\lambda \in \Lambda} \; \mathbb{E}\!\left[ \log \frac{\| R_{\mathrm{roll}} - \alpha(\lambda) \odot W_{\lambda} R_{\mathrm{sync}} \|_F^2 + \epsilon}{\| R_{\mathrm{roll}} \|_F^2 + \epsilon} \right],
\label{eq:lambdastar}
\end{equation}
over a candidate grid $\Lambda = \{\lambda_1, \ldots, \lambda_M\}$. Here, $\lambda$ controls temporal smoothness and $\alpha_l$ controls correction strength at position $l$. We calibrate one pair $(\lambda^*,\alpha)$ for each gap length $S$; the same gap-specific pair is shared by all windows and by both velocity components. Validation is used for model selection, not as an inference-time gate. It can discourage degradation on average over held-out windows, but it does not protect any individual test interval. The temporally smoothed field is
\begin{equation}
X^{W}_{\ell} = \hat{X}^{-}_{\ell} + \alpha_{\ell} (W_{\lambda^*} R_{\mathrm{sync}})_{\ell}.
\label{eq:wrepair}
\end{equation}

\textbf{Energy target estimation.} After temporal smoothing, we estimate the desired energy trajectory from synchronized frames on both sides of the interval. No reference field from inside the gap is used at test-time inference. With
\begin{equation}
E(X_t) = \frac{1}{|\Omega|} \sum_{x,y} \tfrac{1}{2} X_t(x,y)^2,
\label{eq:energydef}
\end{equation}
let $\mathcal{T}_{\mathrm{sync}}$ contain the timestamps of the leading and trailing synchronized frames used for one repair interval, and let $E_t$ be the corresponding energy computed from their frozen-VAE reconstructions. 
For an interval of length $L$, $\mathcal{T}_{\mathrm{sync}}=\{-3,-2,-1,0,L+1,L+2,L+3,L+4\}$, while the repaired positions are indexed by $t\in\{1,\ldots,L\}$.
We fit the cubic polynomial
\begin{equation}
 p(t;\mathbf{a}) = a_0 + a_1 t + a_2 t^2 + a_3 t^3
 \label{eq:cubicpoly}
\end{equation}
by least squares in log-energy space,
\begin{equation}
 \mathbf{a}^{*} = \operatorname*{argmin}_{\mathbf{a}\in\mathbb{R}^{4}}
 \sum_{t\in\mathcal{T}_{\mathrm{sync}}}
 \left[p(t;\mathbf{a})-\log\!\left(\max(E_t,\epsilon_E)\right)\right]^2,
 \label{eq:cubicfit}
\end{equation}
and evaluate it inside the gap to obtain the positive target
\begin{equation}
 \hat{E}(t) = \exp\!\left(p(t;\mathbf{a}^{*})\right).
\label{eq:etarget}
\end{equation}
The same regression is applied independently to each $16\times16$ region to obtain $\hat{E}_{mn}(t)$. This is a cubic regression through all selected synchronized energy samples, not a cubic determined from only two scalar endpoint values. Fitting in log space stabilizes relative variation, while exponentiation keeps every target positive.
We use $\epsilon_E=10^{-12}$, with no normalization of the time coordinates and no additional clipping or overshoot guard on the fitted trajectory.

\textbf{Analytic regional correction.} For compactness, define $[E]_{\epsilon_E}=\max(E,\epsilon_E)$. The temporally smoothed field already contains the predicted spatial structure, so we adjust only its regional amplitude. The gain follows directly from the mismatch between its current and target regional energies:
\begin{equation}
g_{mn}(t)=
\sqrt{\frac{[\hat{E}_{mn}(t)]_{\epsilon_E}}
{[E_{mn}(X^{W}(t))]_{\epsilon_E}}}.
\label{eq:gain}
\end{equation}

\textbf{Smooth spatial modulation.} Scaling each block independently would create artificial discontinuities. Each $16\times16$-pixel region contributes one gain, giving a $16\times16$ coarse gain map for the $256\times256$ field. We represent these gains in log space, bilinearly upsample the coarse log-gain map, and exponentiate the result to obtain a smooth spatial gain field $g(x,y,t)$. We then apply $\widetilde{X}(x,y,t) = g(x,y,t)\, X^{W}(x,y,t)$. Because upsampling blends neighboring gains, the resulting regional energies move toward, but do not exactly match, all coarse regional targets. The lower bound $[\cdot]_{\epsilon_E}$ handles zero or near-zero energies; no additional clipping is applied to either the regional or global gains.

\textbf{Global energy lock.} Regional modulation adjusts the spatial energy distribution but does not guarantee the global target. A final frame-wise scalar removes the remaining mismatch:
\begin{equation}
g_{\mathrm{glob}}(t)=
\sqrt{\frac{[\hat{E}_{\mathrm{glob}}(t)]_{\epsilon_E}}
{[E_{\mathrm{glob}}(\widetilde{X}(t))]_{\epsilon_E}}},
\qquad
\hat{X}^{+}_t=g_{\mathrm{glob}}(t)\widetilde{X}(t).
\label{eq:globallock}
\end{equation}
The final field matches the safeguarded boundary-inferred global target $\hat{E}_{\mathrm{glob}}(t)$ up to numerical precision. Reported global-energy error instead measures that target's accuracy against the withheld authoritative energy.

\textbf{Calibration.} The temporal smoother is calibrated over $\lambda\in\{0\}\cup\{10^{-4+0.5k}\}_{k=0}^{20}$ using six overlapping windows sampled at a four-frame stride, with four used for fitting and two held out for validation. Lag-wise $\alpha$ and $\lambda^{*}$ are selected as in Eqs.~\eqref{eq:alpha}--\eqref{eq:lambdastar}, with a numerical floor of $10^{-12}$ in the ratios; held-out validation sets $\alpha$ to zero for any lag that does not reduce residual error. Calibration samples from both velocity components are pooled so that one gap-specific $(\lambda^*,\alpha)$ is used for both $v_x$ and $v_y$.

\textbf{Inference.} Test-time inference uses only the retained leading and trailing latent blocks, their frozen-VAE reconstructions, the decoded rollout, and the precomputed $(W_{\lambda^*},\alpha)$. The boundary reconstructions provide the energy values in Eq.~\eqref{eq:etarget}. Given those values, Eqs.~\eqref{eq:etarget}--\eqref{eq:globallock} are closed form and require no additional learned correction network. The original full-resolution $X^{\mathrm{P}}$ fields are unavailable to the repair method; hidden fields inside the gap are used only for evaluation.

\subsection{Operational Semantics}
Before resynchronization, the twin can expose only $\hat{X}^{-}_{\ell}$; Eq.~\eqref{eq:globallock} is not yet available. The current method uses all $K$ states in the trailing synchronization block. If that block ends at $\tau_j + S + K$, the reconciled version of position $\ell$ becomes available after $S + K - \ell$ time steps. Systems using this method should therefore label version and provenance explicitly.

Reconciliation does not change the latent state that seeds the next autonomous interval: the next rollout begins directly from the newly received primary-derived $K$-state latent block. This boundary carries no operational penalty under block synchronization: reconciliation completes precisely when the next authoritative block arrives, and re-seeding from that primary-derived block strictly dominates any repaired latent built from the twin's own outputs. Corrected-state feedback could matter only during extended outages spanning multiple gaps without resynchronization; latent-space reconciliation targets that case and remains future work.

\subsection{Representation Boundary}
The causal stage advances a reduced latent state and decodes a field when a physical output or QoI is needed. The retrospective stage uses decoded fields, one application of the $L \times K$ bridge, regional energy evaluations, one bilinear upsampling of the gain map, and two multiplicative gains per reconciled state; it uses no Fourier transform or additional correction network and introduces no trained parameters. Runtime and peak-memory measurements are not available, so we do not make system-level efficiency claims or infer deployment savings from latent dimensionality alone.

\section{Experiments}
\label{sec:experiments}

\subsection{Setup}
\label{sec:setup}
\textbf{Dataset.} We use JHTDB forced isotropic turbulence (\texttt{isotropic1024coarse})~\cite{jhtdb}. Sixteen $512\times512$ $xy$ slices at $z$ indices $\{1,17,33,\ldots,241\}$ are partitioned into four non-overlapping $256\times256$ patches, yielding 64 spatial sequences. They share 256 timestamps at $\Delta t=0.002$; four patches belong to each slice and are not independent realizations. Each field is normalized to zero mean and unit range before encoding. Frames 1--160, 161--192, and 193--256 (1-indexed) are used for training, validation, and testing. Thus, testing is held out in time but not by slice. Both $v_x$ and $v_y$ use this protocol.

\textbf{Synchronization-loss emulation.} The predictor consumes four latent states and produces the next four. For longer horizons, its predictions are fed back recursively. We evaluate retrospective reconciliation on overlapping test windows with a four-frame stride. For every window, the repaired interval spans the full synchronization gap, i.e., $L=S$, and is immediately followed by $K=4$ synchronized states. In 1-indexed coordinates, the repair-window starts are $193+4j$, with $j=0,\ldots,14$ for $S=4$ and $j=0,\ldots,13$ for $S\in\{6,8\}$, giving 15, 14, and 14 windows, respectively. We also apply the repeating block schedule from Section~\ref{sec:formulation} to test states 193--256. For each $S$, this schedule is fixed a priori and is not adjusted based on test errors.

The gap-specific $(\lambda^*,\alpha)$ calibration of Section~\ref{sec:energyproj} uses no test window or hidden physical reference.

\textbf{Metrics.} All reported metrics compare decoded predictions with the original physical reference $X^{\mathrm{P}}$, not with $X^{\mathrm{VAE}}$. For a velocity field $v$ and reconstruction $\hat{v}$, we use range-normalized RMSE,
\begin{equation}
\mathrm{NRMSE}(\hat{v},v)=
\frac{\sqrt{\langle(\hat{v}-v)^2\rangle_{\Omega}}}
{\max(v_{\max}-v_{\min},\epsilon_q)},
\label{eq:nrmse}
\end{equation}
where $\langle\cdot\rangle_{\Omega}$ is the spatial mean and $v_{\max}$ and $v_{\min}$ are computed from the reference field independently for each patch sequence and frame. We also report relative errors for global component energy $E^G$, regional energy $E^R$ on $16\times16$ blocks, global longitudinal-gradient intensity $D^G_L$, and the corresponding regional map $D^R_L$.

For a scalar global QoI $q^G$ and a vectorized regional QoI map $q^R$, the reported relative errors are
\begin{align}
e_G &= \frac{|q^G(\hat v)-q^G(v)|}{\max(|q^G(v)|,\epsilon_q)}, &
e_R &= \frac{\|q^R(\hat v)-q^R(v)\|_2}{\max(\|q^R(v)\|_2,\epsilon_q)},
\label{eq:qoierrors}
\end{align}
with $\epsilon_q=10^{-12}$. The QoI errors are multiplied by 100 when reported as percentages. All metrics are first computed per patch sequence and frame. Results at each rollout position are then averaged equally over all patch-sequence--window samples, and scalar table entries are further averaged over rollout positions.

\textbf{Comparators and variants.} All methods use the same synchronization schedule and evaluation intervals. \emph{Provisional rollout} is the causal output. \emph{Physical linear interpolation} is a future-aware comparator constructed from synchronized decoded boundaries. Specifically, for gap position $\ell\in\{1,\ldots,S\}$,
\[
X^{\mathrm{lin}}_{\tau_j+\ell}
=
\left(1-\frac{\ell}{S+1}\right)X^{\mathrm{VAE}}_{\tau_j}
+
\frac{\ell}{S+1}X^{\mathrm{VAE}}_{\tau_j+S+1}.
\]
The implementation therefore interpolates between the last decoded state of the leading synchronization block and the first decoded state of the trailing synchronization block. The \emph{temporal smoother} applies the calibrated bridge $\alpha W_{\lambda^*}$ in Eq.~\eqref{eq:wrepair}; \emph{full reconciliation} then adds analytic energy matching.

The evaluation addresses three questions: (RQ1) how rapidly does provisional QoI fidelity degrade with time since synchronization; (RQ2) how much of that degradation is removed after resynchronization; and (RQ3) how do the two outputs change with synchronization gap $S$, including comparison with a future-aware interpolation baseline?

RQ1 and RQ2 deliberately read the same $S=8$ position-wise experiment in two ways: RQ1 examines the causal provisional trajectory, whereas RQ2 measures its later future-conditioned revision.

\subsection{RQ1: Provisional QoI Drift Within a Synchronization Gap}
\label{sec:rq1}
\begin{figure*}[!t]
\centering
\begin{subfigure}{0.32\textwidth}
\includegraphics[width=\textwidth]{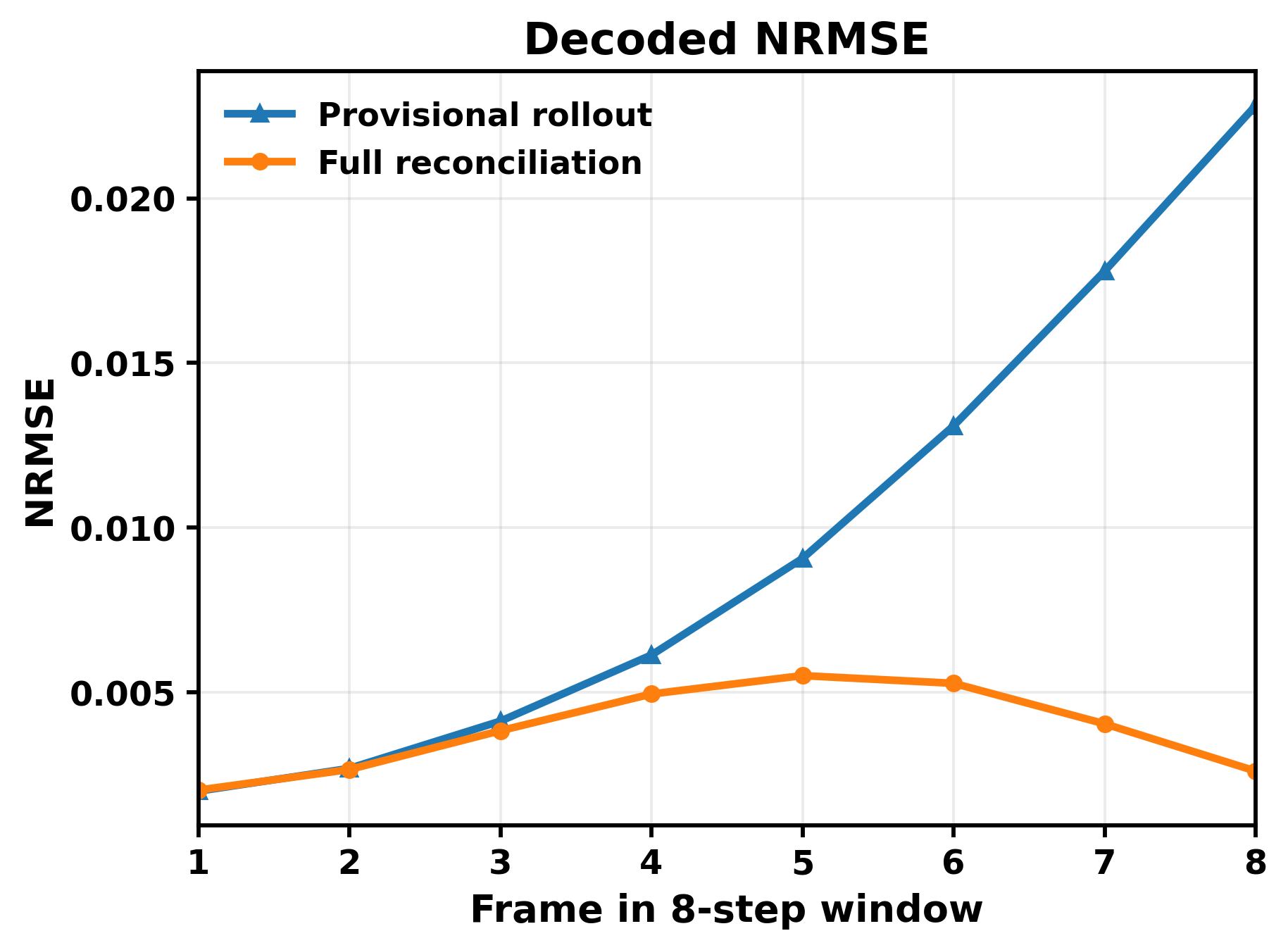}
\caption{Decoded NRMSE}
\end{subfigure}
\hfill
\begin{subfigure}{0.32\textwidth}
\includegraphics[width=\textwidth]{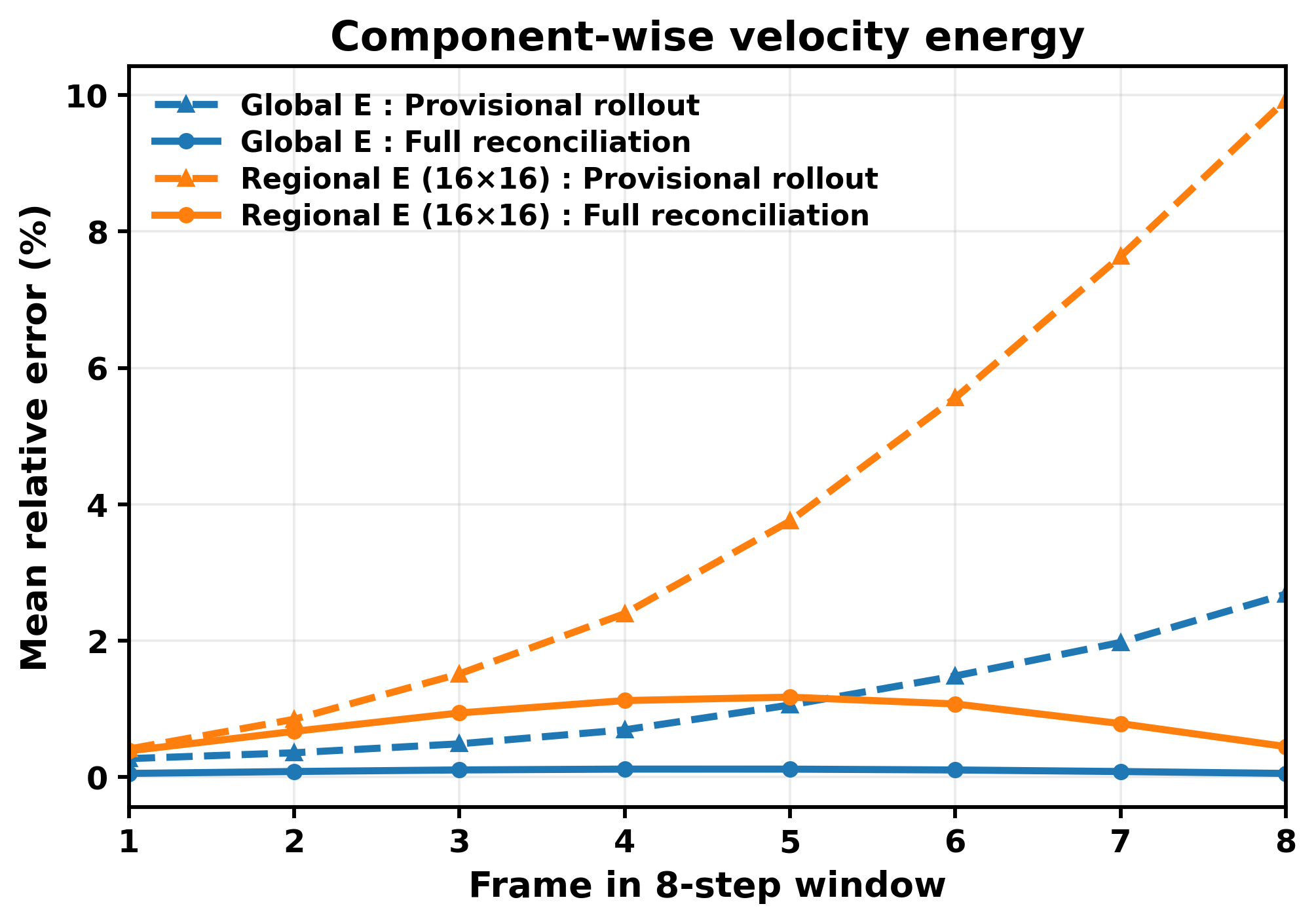}
\caption{Component-energy error}
\end{subfigure}
\hfill
\begin{subfigure}{0.32\textwidth}
\includegraphics[width=\textwidth]{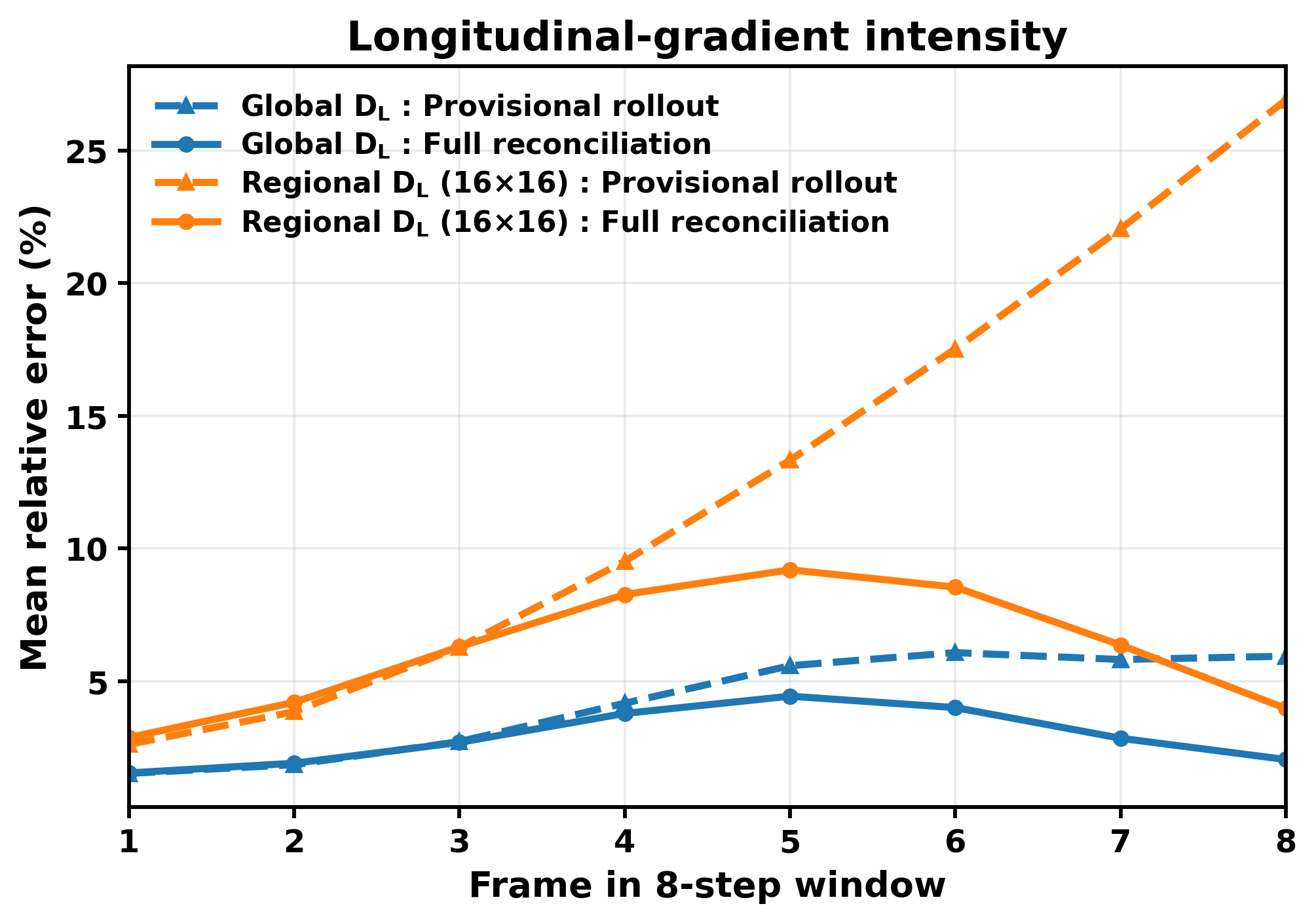}
\caption{Longitudinal-gradient-intensity error}
\end{subfigure}
\caption{Position-wise provisional and full-reconciliation errors over 14 overlapping eight-state intervals, averaged across 64 spatial patch sequences ($v_x$). The provisional curves show QoI drift during the longest tested gap (RQ1); comparison with the reconciled curves shows the recovery available after the following four synchronized states arrive (RQ2). Full reconciliation is the calibrated temporal smoother $\alpha W_{\lambda^*}$ followed by energy matching. Panels (b) and (c) include global and $16\times16$ regional QoIs. Lower is better.}
\label{fig:repair}
\end{figure*}

Fig.~\ref{fig:repair} reports the current evaluation round for the longest gap, $S=8$, over 14 overlapping test intervals and 64 spatial patch sequences. Focusing first on the provisional curves, decoded NRMSE rises from approximately 0.002 at the first position to 0.023 at the eighth. Global component-energy error remains below 3\%, while regional component-energy error reaches approximately 10\%. The global and regional longitudinal-gradient-intensity errors reach approximately 6\% and 27\%, respectively. Because NRMSE and the QoI errors use different normalizations, their numerical ratio has no direct physical meaning; the important result is their distinct position-wise growth, which motivates evaluating each provisional interval through the QoIs required by its downstream use.

\subsection{RQ2: Retrospective Reconciliation}
\label{sec:rq2}
Fig.~\ref{fig:repair} also overlays the reconciled trajectories. Provisional error grows with position in the interval. Reconciled error first rises and then falls as the trajectory approaches the trailing synchronization boundary, where the observed residual provides the strongest information. This is a pattern in the mean curves, not a bound or monotonicity guarantee for individual intervals.

Table~\ref{tab:framewise} reports the frame-wise reduction between the two mean curves. The benefit is strongly position dependent: at position 1, NRMSE and both gradient metrics are slightly worse after reconciliation, and the regional gradient deficit persists through position 3, whereas at position 8 all five errors fall by 65.53--98.16\%. The early-position gradient deficits are consistent with the energy-matching stage acting where temporal drift is still negligible, so its amplitude adjustment has little drift to offset; the same mechanism appears as the small regional-gradient exception in Section~\ref{sec:energyeffect}. In total, reconciliation improves 34 of the 40 frame--metric entries, with the largest gains near resynchronization, where the trailing boundary constrains accumulated drift most directly.

\begin{table}[!t]
\centering
\caption{Frame-wise error reduction (\%) from provisional rollout to full reconciliation at each position of the eight-state interval ($v_x$). For metric $m$ and position $\ell$, $\bar e^-_{m,\ell}$ and $\bar e^+_{m,\ell}$ are first averaged over 64 spatial patch sequences and 14 overlapping test intervals; each entry is then $100(\bar e^-_{m,\ell}-\bar e^+_{m,\ell})/\bar e^-_{m,\ell}$. Positive values favor reconciliation. The final row is the unweighted mean of the eight position-wise percentages.}
\label{tab:framewise}
\scriptsize
\setlength{\tabcolsep}{4pt}
\begin{tabular}{cccccc}
\toprule
\textbf{Frame} & \textbf{NRMSE} & \textbf{Global $E$} & \textbf{Regional $E$} & \textbf{Global $D_L$} & \textbf{Regional $D_L$} \\
\midrule
1 & -1.51 & 81.53 & 6.82  & -1.19 & -9.93 \\
2 & 1.19  & 77.93 & 21.01 & -2.82 & -9.13 \\
3 & 7.03  & 79.22 & 37.99 & 1.05  & -0.22 \\
4 & 19.31 & 83.62 & 53.27 & 9.24  & 13.16 \\
5 & 39.27 & 89.33 & 68.85 & 20.56 & 31.00 \\
6 & 59.73 & 93.24 & 80.78 & 34.08 & 51.23 \\
7 & 77.28 & 96.08 & 89.78 & 51.00 & 71.14 \\
8 & 88.58 & 98.16 & 95.54 & 65.53 & 85.23 \\
\midrule
\textbf{Position mean} & \textbf{36.36} & \textbf{87.39} & \textbf{56.75} & \textbf{22.18} & \textbf{29.06} \\
\bottomrule
\end{tabular}
\end{table}

\subsection{RQ3: Fidelity vs.\ Synchronization Gap}
\label{sec:rq3}
\begin{figure*}[!t]
\centering
\includegraphics[width=0.88\textwidth]{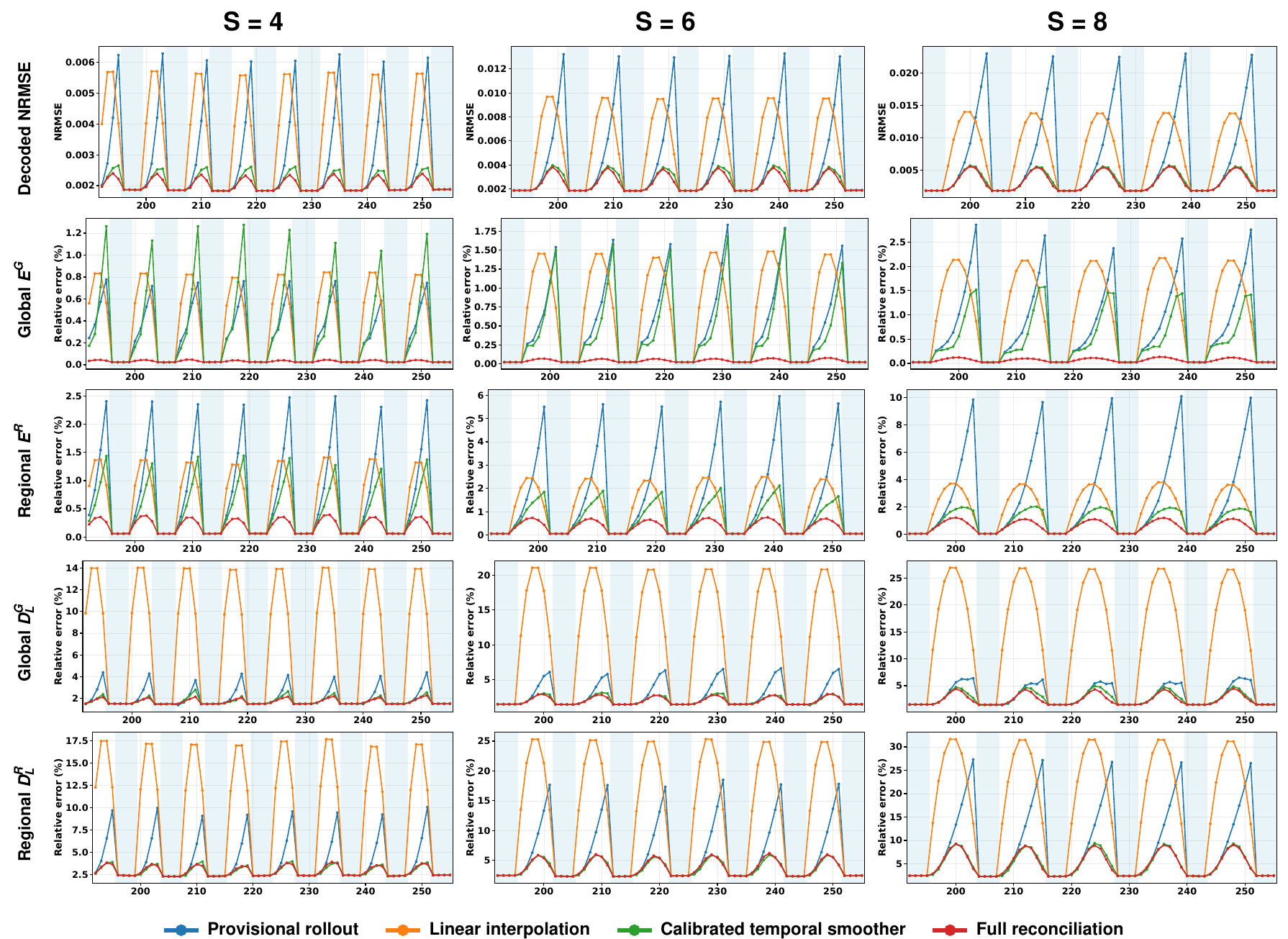}
\caption{Complete $v_x$ error trajectories across the repeating block-synchronization schedules, mean over 64 spatial patch sequences from 16 $z$-slices. Columns show $S=4,6,8$; rows, from top to bottom, show decoded NRMSE, global component-energy error, regional component-energy-map error, global longitudinal-gradient-intensity error, and regional longitudinal-gradient-intensity-map error. Shaded bands are synchronized primary-derived blocks. The shared legend identifies provisional rollout, physical linear interpolation, the calibrated temporal smoother, and full reconciliation.}
\label{fig:syncvx}
\end{figure*}

Fig.~\ref{fig:syncvx} shows the complete repeating block schedules for $v_x$ and all three gaps; the corresponding $v_y$ trajectory atlas is provided in the supplementary material and mirrors the $v_x$ pattern. Shaded regions are primary-derived synchronization blocks. In each unshaded interval, the surrogate first emits provisional states and later replaces them with reconciled states. Error grows after synchronization and resets when primary states return. The figures expose the full temporal pattern; Tables~\ref{tab:meansvx} and~\ref{tab:meansvy} provide the corresponding window-averaged means.

\begin{table*}[!t]
\centering
\caption{Window-averaged mean errors for $v_x$ over 64 spatial patch sequences ($\times$15 test windows for $S=4$; $\times$14 for $S=6,8$). The patches come from 16 $z$-slices and share timestamps; test windows overlap. Energy and gradient-intensity errors are percentages; NRMSE is dimensionless. Bold marks the numerical minimum in each gap--metric block; it does not imply statistical significance.}
\label{tab:meansvx}
\scriptsize
\setlength{\tabcolsep}{4pt}
\begin{tabular}{clccccc}
\toprule
\textbf{Gap} & \textbf{Method} & \textbf{NRMSE} & \textbf{Global $E$ (\%)} & \textbf{Regional $E$ (\%)} & \textbf{Global $D_L$ (\%)} & \textbf{Regional $D_L$ (\%)} \\
\midrule
\multirow{4}{*}{$S=4$}
& Provisional rollout      & 0.003736 & 0.4481 & 1.2902 & 2.5766  & 5.5791 \\
& Physical interpolation   & 0.004809 & 0.6936 & 1.1232 & 11.8651 & 14.6462 \\
& Temporal smoother        & 0.002319 & 0.5790 & 0.7769 & 1.9337  & \textbf{3.2866} \\
& Full reconciliation      & \textbf{0.002184} & \textbf{0.0388} & \textbf{0.2992} & \textbf{1.8597} & 3.3323 \\
\midrule
\multirow{4}{*}{$S=6$}
& Provisional rollout      & 0.006190 & 0.7596 & 2.4312 & 3.6705  & 8.8646 \\
& Physical interpolation   & 0.007489 & 1.1371 & 1.9008 & 16.6034 & 19.9101 \\
& Temporal smoother        & 0.003063 & 0.6252 & 1.0955 & 2.3996  & \textbf{4.4767} \\
& Full reconciliation      & \textbf{0.002912} & \textbf{0.0587} & \textbf{0.5394} & \textbf{2.3133} & 4.5908 \\
\midrule
\multirow{4}{*}{$S=8$}
& Provisional rollout      & 0.009713 & 1.1220 & 4.0038 & 4.2134  & 12.7628 \\
& Physical interpolation   & 0.010365 & 1.6057 & 2.7728 & 20.4007 & 24.0433 \\
& Temporal smoother        & 0.004012 & 0.7232 & 1.4344 & 3.2328  & \textbf{6.1970} \\
& Full reconciliation      & \textbf{0.003861} & \textbf{0.0849} & \textbf{0.8202} & \textbf{2.9101} & 6.2175 \\
\bottomrule
\end{tabular}
\end{table*}

\begin{table*}[!t]
\centering
\caption{Window-averaged mean errors for $v_y$ under the same protocol as Table~\ref{tab:meansvx}. Both components show comparable error magnitudes. Bold marks the numerical minimum within each gap and metric, as in Table~\ref{tab:meansvx}.}
\label{tab:meansvy}
\scriptsize
\setlength{\tabcolsep}{4pt}
\begin{tabular}{clccccc}
\toprule
\textbf{Gap} & \textbf{Method} & \textbf{NRMSE} & \textbf{Global $E$ (\%)} & \textbf{Regional $E$ (\%)} & \textbf{Global $D_L$ (\%)} & \textbf{Regional $D_L$ (\%)} \\
\midrule
\multirow{4}{*}{$S=4$}
& Provisional rollout      & 0.003292 & 0.1889 & 0.6422 & 2.7068  & 5.7360 \\
& Physical interpolation   & 0.003846 & 0.1929 & 0.4558 & 12.9086 & 15.9107 \\
& Temporal smoother        & 0.002192 & 0.3052 & 0.4650 & 2.1671  & \textbf{3.6582} \\
& Full reconciliation      & \textbf{0.002026} & \textbf{0.0183} & \textbf{0.1413} & \textbf{2.0856} & 3.7232 \\
\midrule
\multirow{4}{*}{$S=6$}
& Provisional rollout      & 0.005401 & 0.3236 & 1.2180 & 4.0387  & 9.1164 \\
& Physical interpolation   & 0.005920 & 0.3125 & 0.7885 & 18.0514 & 21.5389 \\
& Temporal smoother        & 0.002834 & 0.3775 & 0.6674 & 2.5723  & \textbf{4.7926} \\
& Full reconciliation      & \textbf{0.002606} & \textbf{0.0242} & \textbf{0.2516} & \textbf{2.4758} & 4.9369 \\
\midrule
\multirow{4}{*}{$S=8$}
& Provisional rollout      & 0.008460 & 0.5567 & 2.0687 & 4.6846  & 13.0058 \\
& Physical interpolation   & 0.008184 & 0.4395 & 1.1750 & 22.0598 & 25.7901 \\
& Temporal smoother        & 0.003565 & 0.3995 & 0.7937 & 3.0141  & \textbf{6.2256} \\
& Full reconciliation      & \textbf{0.003355} & \textbf{0.0329} & \textbf{0.3650} & \textbf{2.8246} & 6.3741 \\
\bottomrule
\end{tabular}
\end{table*}

For $v_x$, errors increase with the gap. Full reconciliation has the lowest mean NRMSE, energy errors, and global gradient error at every gap; the smoother is lower on regional gradient error by at most 0.12 percentage points. At $S=8$, reconciliation's NRMSE is 0.0039 versus 0.0097 for provisional rollout and 0.0104 for interpolation; global gradient error is 2.91\% versus 4.21\% and 20.40\%. Interpolation's global gradient error rises from 11.87\% to 20.40\% as $S$ grows from 4 to 8, while reconciliation remains at 1.86--2.91\%. Endpoint blending may attenuate fine-scale structure that the latent trajectory retains. These are patch/window means, not per-patch dominance or independent replication.

The $v_y$ results mirror $v_x$ in both magnitude and pattern. Relative to provisional rollout as $S$ grows from 4 to 8, full reconciliation reduces NRMSE by 38.5--60.3\%, global-energy error by 90.3--94.1\%, regional-energy error by 78.0--82.4\%, global gradient-intensity error by 22.9--39.7\%, and regional gradient-intensity error by 35.1--51.0\%; each reduction increases monotonically with the gap.

As for $v_x$, the temporal smoother is marginally lower than full reconciliation on $v_y$ regional gradient intensity (by at most 0.15 percentage points); full reconciliation is lower on every other metric. These differences are numerical comparisons of means, not evidence that either variant is statistically superior. Physical interpolation again loses fine-scale fidelity decisively: at $S=8$, its global gradient error is 22.06\%, compared with 4.68\% for provisional rollout and 2.82\% after full reconciliation.

\subsection{Effect of Energy Matching}
\label{sec:energyeffect}
\label{sec:ablationE}
Tables~\ref{tab:meansvx}--\ref{tab:meansvy} separate the stages. At $S=8$ for $v_x$, smoothing reduces NRMSE from 0.0097 to 0.0040 and global gradient error from 4.21\% to 3.23\%, but can overshoot energy. Energy matching lowers the smoother's authoritative-reference global-energy error by 93.3\%, 90.6\%, and 88.3\% at $S=4,6,8$ for $v_x$, with similar $v_y$ reductions; this tests the boundary-derived target, not merely constraint satisfaction. In absolute terms, these are reductions from already sub-percent errors (at most 0.72\%) to 0.02--0.08\%: the temporal smoother is the dominant accuracy mechanism, and the practical value of energy matching lies in guaranteeing an internally consistent global-energy trajectory for archived history, together with its regional-energy improvements (e.g., 1.43\% to 0.82\% at $S=8$ for $v_x$). It also improves NRMSE and global gradient error. Regional gradient error is the exception: at $S=8$ for $v_x$, it changes from 6.197\% to 6.217\% and increases for 37 of 64 patches.

Metrics are computed per patch and frame, then averaged equally over patches, windows, and rollout positions. At $S=8$ for $v_x$, paired patch bootstrapping gives reconciliation improvements of $0.00585$ NRMSE (95\% interval $[0.00557,0.00615]$) and 1.30 global-gradient percentage points ($[1.08,1.53]$); only 2 of 64 patches degrade on global gradient and none on the other metrics. These descriptive intervals do not account for within-slice clustering or shared timestamps.

To account for spatial dependence, we additionally perform 10,000 paired cluster-bootstrap resamples over the 16 physical $z$-slices. For the energy-matching comparison at $S=8$ for $v_x$, the 95\% interval for the global-energy improvement is $[0.559,\,0.716]$ percentage points, while the corresponding intervals for NRMSE, regional energy, and global gradient error also exclude zero. Regional gradient remains the only exception: its mean change is an increase of 0.0205 percentage points, with a 95\% interval of $[-0.071,\,0.114]$, and 6 of 16 slice-level averages increase.

\subsection{Discussion and Scope}
The main systems result is the separation of an online product from a revised historical product: provisional fields provide causal continuity, while reconciled fields are delayed finite-interval estimates conditioned on the complete trailing block, serving buffered analysis, archival revision, and other consumers that accept versioned history.

Reduced state is an input assumption, not an evaluated systems benefit: without runtime, memory, hardware, and energy measurements for primary and surrogate, the ratio $N/d$ cannot be read as a deployment advantage.

To test whether the observed gains are specific to the original test segment, we additionally evaluate the frozen models at $S=8$ under two 64-frame holdout settings without retraining or recalibration. The \emph{spatial-holdout} setting keeps the original test interval, frames 193--256, but replaces the original 16 $z$-slices with 16 previously unseen slices. The \emph{temporal-holdout} setting keeps the original $z$-slices but evaluates the subsequent 64 frames, 257--320. Table~\ref{tab:ood} shows that full reconciliation retains comparable improvements for both $v_x$ and $v_y$ under both settings, with approximately $60\%$ NRMSE reduction and similarly consistent reductions in the energy and longitudinal-gradient-intensity QoIs. These results indicate that the reconciliation benefit is not confined to the spatial slices or temporal interval used in the original evaluation.

\begin{table}[!t]
\centering
\caption{Full-reconciliation errors on the original test set and the frozen-model spatial and temporal holdout sets at $S=8$. Energy and gradient-intensity errors are percentages; NRMSE is dimensionless.}
\label{tab:ood}
\scriptsize
\setlength{\tabcolsep}{4pt}

\begin{tabular}{clccccc}
\toprule
 & \textbf{Test region} & \textbf{NRMSE} & \textbf{$E^G$ (\%)} & \textbf{$E^R$ (\%)} & \textbf{$D_L^G$ (\%)} & \textbf{$D_L^R$ (\%)} \\
\midrule
\multirow{3}{*}{$v_x$}
& Original      & 0.00386 & 0.0849 & 0.8202 & 2.9101 & 6.2175 \\
& Spat. holdout & 0.00384 & 0.0829 & 0.8049 & 2.8881 & 6.1292 \\
& Temp. holdout & 0.00393 & 0.0856 & 0.8491 & 2.9952 & 6.4860 \\
\midrule
\multirow{3}{*}{$v_y$}
& Original      & 0.00336 & 0.0329 & 0.3650 & 2.8246 & 6.3741 \\
& Spat. holdout & 0.00335 & 0.0319 & 0.3652 & 2.8453 & 6.3264 \\
& Temp. holdout & 0.00335 & 0.0325 & 0.3628 & 2.9900 & 6.6090 \\
\bottomrule
\end{tabular}
\end{table}

Within this regime, gradient-sensitive QoIs grow differently from NRMSE, and reconciliation improves provisional history most near resynchronization. Repair targets $X^{\mathrm{VAE}}$ temporal drift; representation error remains in all reported end-to-end errors. The primary evaluation still contains shared timestamps and overlapping windows, and the additional frozen-model tests remain within the same JHTDB turbulence dataset. Broader future-conditioned reconstruction methods are outside this comparison.

\section{Conclusion}
\label{sec:conclusion}
We studied a block-synchronized reduced-state surrogate that produces two trajectories. When primary synchronization is lost, a causal latent predictor supplies an immediate provisional trajectory. After the complete trailing block arrives, a boundary-anchored physical-space smoother and analytic energy-matching step revise that finite history. The distinction makes the latency and capability of each product clear: provisional continuity is available online, while improved historical fidelity is delayed and future conditioned.

The turbulence study shows why QoI-aware evaluation matters: NRMSE and gradient-sensitive QoIs exhibit distinct position-wise degradation. For both components, full reconciliation gives the lowest window-averaged means on every metric except regional gradient intensity, where the calibrated smoother is marginally lower; separation from interpolation is largest on the gradient QoIs. Energy matching enforces its boundary-inferred global target without an additional correction network; its improvements against the authoritative references are empirical.

Future work should extend the evaluation to other datasets and flow regimes, use cluster-aware uncertainty analysis, and develop latent-space reconciliation for consumers that require a corrected internal state. A separate systems study should measure primary and surrogate runtime, hardware, memory, and replay cost, and should define application-specific QoI tolerances.

\makeatletter\def\@IEEEbibitemsep{0pt plus 0.3pt}\makeatother

\end{document}